\documentclass[11pt,twoside,a4paper]{article}
\usepackage{graphicx,color,amsmath}
\usepackage{xspace}
\usepackage{booktabs}
\usepackage{cite}
\usepackage{epsfig}
\usepackage{feynmp}
\usepackage{a4wide}
\usepackage[justification=centering]{subfig}
\usepackage{pifont}

\newcommand{\tsc}[1]{\textsc{#1}}
\newcommand{\ttt}[1]{\texttt{#1}}

\newcommand{\desai}[1]{\textcolor{blue}{\emph{#1}}}

\newcommand{\TeV}{\,\mbox{Te\kern-0.2exV}}
\newcommand{\GeV}{\,\mbox{Ge\kern-0.2exV}\xspace}
\newcommand{\MeV}{\,\mbox{Me\kern-0.2exV}}
\newcommand{\keV}{\,\mbox{ke\kern-0.2exV}}
\newcommand{\eV}{\,\mbox{e\kern-0.2exV}}

\newcommand{\Hw}{\tsc{Herwig}}

\newcommand{\Mg}{\tsc{MadGraph}}
\newcommand{\Py}{\tsc{Pythia}}

\newcommand{\eqRef}[1]{equation~\eqref{#1}\xspace}
\newcommand{\secRef}[1]{section~\ref{#1}\xspace}
\newcommand{\SecRef}[1]{Section~\ref{#1}\xspace}

\newcommand{\FigRef}[1]{Fig.~\ref{#1}\xspace}

\newcommand{\FigsRef}[1]{Figs.~\ref{#1}\xspace}

\begin{document}
\begin{minipage}{\textwidth}
\flushright
CERN-PH-TH/2011-210\\
MCNET-11-23
\end{minipage}
\vskip5mm
\begin{center}
{\Large Supersymmetry and Generic BSM Models in PYTHIA 8}
\end{center}
\vskip5mm
\begin{center}
{\large Nishita~Desai$^1$, Peter~Z.~Skands$^2$}
\end{center}

\parbox{0.9\textwidth}{
\begin{center}
$^1$Harish-Chandra Research Institute, Allahabad 211019, India.  

$^2$Theoretical Physics, CERN, CH-1211 Geneva 23, Switzerland
\end{center}
}

\vskip5mm
\begin{center}
\parbox{0.85\textwidth}{
\begin{center}
\textbf{Abstract}
\end{center}\small

We describe  the implementation of supersymmetric models in \Py~8
including production and decay of superparticles and allowing for 
violation of flavour, CP, and R-parity. We also present a
framework for importing generic new-physics matrix elements into
\Py~8, in a way suitable for use with automated tools. We emphasize
that this possibility should not be viewed as the only way to
implement new-physics models in \Py~8, but merely as an additional
possibility on top of the already existing ones. Finally we address
parton showers in exotic colour topologies, in particular ones involving
colour-epsilon tensors and colour sextets.
}
\end{center}

\section{Introduction}

The Standard Model (SM) of particle physics has been enormously
successful in describing interactions between fundamental particles.
The only experimentally unverified component of the SM is the Higgs
boson which is thought to underlie the electroweak symmetry breaking
(EWSB).  However, experimental evidence for dark matter, neutrino
masses and the theoretical requirement of naturalness appeals for a
theory beyond standard model (BSM).  Monte Carlo generators fulfil an
important role, both in testing the SM to high precision and in
testing the prediction of new theories, by providing a systematic
procedure of comparing theoretical prediction to experimental
observation.

\Py~8~\cite{Sjostrand:2007gs} is a general-purpose Monte Carlo event
generator \cite{Buckley:2011ms} for a full 
simulation of high-energy collision events.  It includes a
comprehensive library of hard-scattering processes, particle decays,
initial- and final-state parton-shower models \cite{Sjostrand:2004ef,Corke:2010yf}, 
hadronization through
string fragmentation \cite{Andersson:1983ia} 
and models of beam remnants and multiple
interactions \cite{Sjostrand:2004pf,Corke:2011yy}. 
It contains a native implementation of a wide variety
of SM and BSM processes and also provides a standard interface
\cite{Alwall:2006yp,Alwall:2007mw} 
to external programs which may be used by a standalone generator.  

We describe here, the updates to the \Py~8 event generator to include
the popular BSM model of supersymmetry (SUSY), additions made to
parton showers and hadronization algorithms to allow for exotic colour
topologies and generic enhancements made to enable interfacing to
parton-level BSM generators.  In \secRef{sec:susy}, we describe the
implementation of supersymmetric models in \Py~8, including production
and decay of superparticles. In \secRef{sec:generic}, we present a
framework for importing generic new-physics matrix elements into
\Py~8, in a way suitable for use with automated tools. In
\secRef{sec:showers}, we discuss the treatment of parton showers in
exotic colour topologies. \SecRef{sec:conclusions} contains a brief
summary and conclusions.

\section{Supersymmetry in \Py~8 \label{sec:susy}}

Supersymmetry (see \cite{Martin:1997ns} for a pedagogical
introduction) is considered one of the best motivated extensions of
the SM due to its ability to address many outstanding theoretical and
experimental issues.  In particular, the Minimal Supersymmetric
extension of the Standard Model (MSSM) is currently a popular
candidate for a BSM theory. The MSSM extends the SM by the addition of
one pair of SUSY generators which implies the presence of one
superpartner to each SM state. The MSSM particle spectrum therefore
has squarks ($\tilde q_i$), sleptons ($\tilde \ell_i$) and gauginos
($\tilde B$, $\tilde W^i$ and $\tilde g$) as the supersymmetric
counterparts of quarks, leptons and gauge bosons respectively.  The
requirement of self-consistency of the theory via anomaly cancellation
also demands two Higgs doublet fields $H_u$ and $H_d$.  After
electroweak symmetry breaking (EWSB), we are left with five Higgs
degrees of freedom viz.\ the CP-even $h_0$ and $H_0$, the CP-odd $A_0$
and two charged Higgs bosons $H^{\pm}$.  The superpartners of the
Higgses --- the fermionic ``Higgsinos'' --- mix with the gauginos to
form neutralinos and charginos.  In particular, the neutral Higgsinos
($\tilde H_1$ and $\tilde H_2$) mix with the neutral $U(1)$ and
$SU(2)$ gauginos ($\tilde B$ and $\tilde W^3$) to form the mass
eigenstates called the neutralinos ($\tilde \chi_i^0;i=1-4$.)
Similarly, the charged Higgsino mixes with the charged $SU(2)$ gaugino
to form charginos ($\tilde \chi_i^\pm; i=1,2$.) The next-to-minimal
supersymmetric extension of the SM (nMSSM) extends this scenario by
adding one extra singlet Higgs field.  This adds another member to the
neutralinos and the neutralino mixing matrix is enlarged to $5 \times
5$.  The current implementation of \Py~8 includes the nMSSM extension
and allows processes with CP, flavour or R-parity violation.

\Py~8 uses the standard PDG codes for numbering the superpartners
\cite{pdg2012} and the particle spectrum is read in via an
SLHA file~\cite{Skands:2003cj,Allanach:2008qq}.  We use the super-CKM
basis (in the conventions of the SLHA2~\cite{Allanach:2008qq}) for
describing the squark sector which allows non-minimal flavour
violation.  The mass-eigenstates of the squarks are then related to
the left- and right-handed squarks via a $6 \times 6$ complex mixing
matrix.  Our implementation can therefore be used to study both CP
violation and flavour violation in the squark sector.
\begin{eqnarray}
\left( \begin{array}{l}
\tilde u_1 \\ \tilde  u_2 \\ \tilde u_3 \\ \tilde u_4 \\ \tilde u_5 \\ \tilde u_6 \end{array} \right) = 
R^u \left( \begin{array}{l}
\tilde u_L \\ \tilde c_L \\ \tilde t_L \\ \tilde u_R \\ \tilde c_R \\ \tilde t_R \end{array} \right); 
\left( \begin{array}{l}
\tilde d_1 \\ \tilde d_2 \\ \tilde d_3 \\ \tilde d_4 \\ \tilde d_5 \\ \tilde d_6 \end{array} \right) = 
R^d \left( \begin{array}{l}
\tilde d_L \\ \tilde s_L \\ \tilde b_L \\ \tilde d_R \\ \tilde s_R \\ \tilde b_R \end{array} \right)
\end{eqnarray}

The neutralino mixing matrix $\mathcal{N}$ is a $4\times 4$ ($5 \times
5$ in the case of nMSSM) mixing matrix describing the transformation
of the gauge eigenstate fermions ($-i \tilde B, -i \tilde W_3, H_1,
H_2$) into the mass eigenstates ($\tilde \chi_1^0, \tilde \chi_2^0,
\tilde \chi_3^0, \tilde \chi_4^0$).  The two chargino mixing matrices
$\mathcal{U}$ and $\mathcal{V}$ describe the diagonalization of the
chargino mass matrix from the gauge eigenstates ($-i W^+,H^+$) to
($\tilde \chi_1^+, \tilde \chi_2^+)$.  Supplementary conventions for 
vertices and most of the cross-section formulae 
are taken from \cite{Bozzi:2007me}, as detailed below. 

\subsection{Couplings}

\Py~8 reads particle masses and mixing matrices via the SUSY Les
Houches Accord (SLHA2) framework \cite{Allanach:2008qq}.  (Read-in of
SLHA1 spectra \cite{Skands:2003cj} is also supported, but mixing the
two standards is strongly discouraged, as the internal translation
from SLHA1 to SLHA2 has only been designed with the original SLHA1 in
mind.)  The raw data read in by the \ttt{SusyLesHouches} class is
accessed by the \ttt{CoupSUSY} class which uses the information to
construct all the SUSY couplings. The couplings are defined according
to \cite{Bozzi:2007me} for all cases except for couplings of
superparticles to Higgs bosons which are defined according to
\cite{Gunion:1987yh}.

The running of electroweak and strong couplings is carried over from
the corresponding one-loop calculations in the Standard Model.  The
\texttt{GAUGE} block can be used to set the boundary values of all
three SM couplings at the SUSY breaking scale.  By default, the masses
of $W$ and $Z$ are assumed to be the pole masses and are used to
calculate the on-shell value of $\sin^2 \theta_W=1-m_W^2/m_Z^2$.  
If externally provided in the 
SLHA file, the value of $\sin \theta_W$ can be set to the running
value using the flag \ttt{SUSY:sin2thetaWMode = 2} (see the \Py~8 HTML
user reference included with the code \cite{Sjostrand:2007gs}). 
The ratio of the two Higgs vacuum expectation values ($\tan \beta$) is
read in from the low scale 
value provided by the \ttt{MINPAR} and \ttt{EXTPAR} blocks.  The default value of
the Higgs mixing angle ($\alpha_H$) is set to the SM limit ($\beta - \pi/2$) which is then overwritten by the
contents of the \ttt{HMIX} block.   

Since the SLHA interface has been extended and can now be used to pass
information on any new particles and decays \cite{Alwall:2007mw}, the
presence of the \ttt{MODSEL} block is used as an indicator of SUSY
models and \Py~8 will initialize the \ttt{CoupSUSY} class only if this
block is present.  Skipping the \ttt{MODSEL} block is acceptable for
Les Houches Event files (LHEF) as long as the user supplies an
external decay table for all required cascade decays.

\subsection{R-parity violation}

The most general MSSM superpotential allows both lepton and
baryon-number violating processes.  This is generally avoided by
demanding invariance under an R-parity defined as $(-1)^{3B-L+2S}$.
From this definition, all SM particles are even whereas all
superpartners are odd under R-parity.  A well known consequence of
this is that the Lightest SUSY particle (LSP) must be stable.  A
neutral, weakly interacting LSP can therefore be a good candidate for
dark matter.  However, the imposition of R-parity can be considered an
aesthetic requirement rather than a consistency requirement and
possible R-parity violating interactions, if present, can be probed by
collider experiments.  We therefore include R-parity violating
production and decay processes in our implementation.

In SLHA conventions, the R-parity violating superpotential is given by
\begin{equation}
\mathcal{W}_{RPV} = \mu_i H_u L_i + \frac{1}{2} \lambda_{ijk} L_i L_j
E_k + \lambda'_{ijk} L_i Q_j D_k + \frac{1}{2}\lambda''_{ijk} U^c_i
D^c_j D^c_k
\end{equation}
The $\mu$-type terms correspond to bi-linear R-parity violation which
causes a mixing between the leptons and neutralinos/charginos.  The
$\lambda$ and  $\lambda'$-type terms lead to lepton number violation
whereas $\lambda''$-type terms lead to baryon-number violation.  The
current implementation does not include the effects of the bi-linear
term.    The R-parity violating couplings $\lambda_{ijk}$ are
antisymmetric under $i \leftrightarrow j$.  Therefore only couplings
for $i>j$ are read and the rest are set by the symmetry property.
Similarly, $\lambda''_{ijk}$ is antisymmetric under $j\leftrightarrow
k$ and hence only couplings with $j>k$ need to be provided.
This implementation includes in particular, the resonant production of
a squark via $\lambda''$-type couplings which can be probed at hadron
collider experiments.  The changes made to showering and hadronization
to account for the non-standard colour structure from such terms will
be explicitly described in section~\ref{sec:showers}. 

\subsection{Cross Sections}

The current implementation of SUSY includes all leading-order (LO) 
$2\to2$ production processes with gluinos, squarks, charginos, and
neutralinos in the final state and also $2\to 1\to 2$ resonant
production of squarks via baryon number 
violating couplings.  All available SUSY processes can be turned on
using \ttt{SUSY:all = on}.  Individual subprocesses can then be
selected based on the final state by setting \ttt{SUSY:idA = {\em
    PDGcode}} and \ttt{SUSY:idB = {\em PDGcode}}.  If only \ttt{idA}
is provided, all processes with that particle in the final state are
turned on. Alternatively, one or more production processes can be
turned on using the string \ttt{SUSY:{\em processname} = on}, again
with \ttt{SUSY:idA} and \ttt{SUSY:idB} providing a further level of
subprocess selection. The available subprocess classes are listed in
Table~\ref{tab:prod}.

\begin{table}
\begin{center}
\begin{tabular}{|l|l|}
\hline
Subprocess class & \ttt{processname}\\
\hline
Chargino and neutralino production & \ttt{qqbar2chi0chi0}, \\
                                   & \ttt{qqbar2chi+-chi0}, \\
                                   & \ttt{qqbar2chi+chi-}.\\
\hline
Gaugino squark production          & \ttt{qg2chi0squark},  \\
                                   & \ttt{qg2chi+-squark}.\\
\hline
Gluino production                  & \ttt{gg2gluinogluino}, \\
                                   & \ttt{qqbar2gluinogluino}. \\
\hline
Squark-gluino production           & \ttt{qg2squarkgluino} \\
\hline
Squark-pair production             & \ttt{gg2squarkantisquark}, \\
                                   & \ttt{qqbar2squarkantisquark} \\
                                   & \ttt{qq2squarksquark} \\
\hline
RPV resonant squark production     & \ttt{qq2antisquark} \\
\hline
\end{tabular}
\caption{\label{tab:prod} List of SUSY production processes.  In all cases, charge conjugate processes are turned on by default.}
\end{center}
\end{table}

The squark-antisquark and squark-squark production processes include
contributions from EW diagrams and their interferences. To estimate
the size of these contributions, and/or for purposes
of comparison to other codes that do not include them, the cross
sections can be restricted to include only the strong-interaction 
contributions, using the following flags:
\begin{itemize} 
  \item \ttt{qqbar2squarkantisquark:onlyQCD = true}.
  \item \ttt{qq2squarksquark:onlyQCD = true}.
\end{itemize}

The baryon number violating coupling $\lambda_{ijk}''$ if present, can
induce resonant squark production via the process $d_j d_k \rightarrow
\tilde u^*_i$ which produces a resonant up-type antisquark or via $u_i
d_j \rightarrow \tilde d^*_k$ or $u_i d_k \rightarrow \tilde d^*_j$
which produce a down-type antisquark.  The expression for an up-type
squark production process is
\begin{eqnarray}
\sigma_{\tilde u^*_i} & = & \frac{ 2 \pi}{3 m_i^2} \sum_{jk} \sum_{i'} |\lambda^{''}_{i'jk} (R^u)_{ii'}|^2
\end{eqnarray}

The expression for down-type squarks is similar, taking into account
the symmetry property $\lambda''_{ijk} = - \lambda''_{ikj}$.  We
implement this production process as \ttt{qq2antisquark} and the
charge conjugate process ($\bar q_i \bar q_j \rightarrow \tilde q_k$)
is included by default.

The supersymmetric Higgs sector is identical in many ways to the
Two-Higgs Doublet Model.  The Higgs production processes have already
been implemented in \Py~8 in the \ttt{SigmaHiggs} class.  The
production of the Higgs bosons can be accessed by including the switch
\ttt{HiggsBSM:all=on}.  For specific Higgs processes, please refer to
the HTML user reference included with the code \cite{Sjostrand:2007gs}.

\subsection{Sparticle Decays \label{sec:sparticle_decays}}

SUSY Particle decays are handled by the class
\ttt{SUSYResonanceWidths}.  The user can choose to read in decay
tables via SLHA or use the decay widths calculated by \Py.  As a
default, \Py\ does not calculate the decay width if a table is
externally supplied. Note, however, that while \Py's internal
treatment can include sophistications such as 
matrix-element-based phase-space weighting and 
running widths, channels read in from an SLHA decay
table will be decayed purely according to phase space, with no
matrix-element weighting. The internal treatment should therefore be
preferable, in most cases, and an option for overriding the automatic
read-in of decay tables is provided, by setting the 
flag \ttt{SLHA:useDecayTable = false}, see sec.~\ref{sec:info}.

The decay of a particular particle may be turned off manually using
the standard \Py~8 structure \ttt{{\em PDGcode}:mayDecay = false} 
or by setting its width to zero in the SLHA decay table. In the
former case, the particle will still be distributed according to a
Breit-Wigner distribution with non-zero width, whereas 
it will always be assigned its pole mass in the latter.

Individual decay modes may be switched on/off using the standard \Py~8
methods, documented in the section on ``The Particle Data Scheme'' in
the program's HTML documentation \cite{Sjostrand:2007gs}.  We discuss
ways to switch modes on/off using SLHA decays tables in
section~\ref{sec:info}.

The internal treatment of 2-body decays is so far restricted to on-shell
particles. A mechanism for effectively generating  
3-body decays via sequences of $1\to 2$ decays involving off-shell
particles is foreseen as an update in the near future (and will be
announced in the \Py~8 update notes). An equivalent mechanism 
is already implemented in \Py~8, e.g., for $h\to ZZ$ decays for light
Higgs bosons.

Currently the following R-parity conserving two-body decays are
implemented:

\begin{itemize}
\item $\tilde g \rightarrow \tilde q_i q_j$
\item $\tilde \chi_i^0 \rightarrow \tilde q_i q_j$, \ $\tilde l_i
  l_j$, \ 
  $\tilde \chi^0_j Z$, \ $\tilde \chi^+_j W^-$
\item $\tilde \chi_i^+ \rightarrow \tilde q_i q_j$, \ $\tilde l_i
  l_j$, \ 
  $\tilde \chi^+_j Z$, \ $\tilde \chi^0_j W^+$
\item $\tilde q_i \rightarrow q_j \tilde \chi_k^0$, \ $q_j \tilde
  \chi_k^+$, \ $\tilde q_j Z$, \ $\tilde q_j W^+$
\end{itemize}

Besides these, we also include two-body R-parity violating decays of
squarks via $\lambda'$ ($\tilde q \rightarrow lq'$) and
$\lambda''$-type couplings ($\tilde q \rightarrow q'q'' $).  We also
include the three-body decays of neutralinos through $\lambda''$-type
couplings via an intermediate squark \cite{Dreiner:1999qz}.  For
certain final states in three body decays, partial decay via
sequential two-body decays may also be kinematically allowed.  In this
case, we demand that only the off-shell components of the matrix
element-squared are allowed to contribute to the three-body decay
width.  Any interferences between the off-shell and on-shell
components are also turned off.  The two-body sequential decays then
proceed as normal.

The Higgs boson running widths are calculated in the associated
classes \ttt{ResonanceH} for CP even ($h_0, H_0$) and the CP odd
($A_0$) Higgses, and \ttt{ResonanceHchg} for charged Higgses($H^\pm$).
By default, the Higgs decay tables are not overwritten even if they
are read via SLHA because \Py~8 performs a more accurate phase space
calculation than the flat weighting that is performed for decay widths
read in via SLHA. The decays of Higgses into SUSY particles will be
included in a future update.

\section{Interfacing Generic BSM Models \label{sec:generic}}

The simplest way of implementing a new model may often be to just
inherit from SM or BSM processes that have already been implemented in
\Py~8, modifying and generalizing them as appropriate, as described in
the section on ``Semi-internal Processes'' in the main \Py~8
documentation \cite{Sjostrand:2007gs}.

Alternatively, \Py~8 can read in parton-level events generated by
external matrix-element event generators
\cite{Pukhov:2004ca,Boos:2004kh,Alwall:2007st,Alwall:2011uj,Kilian:2007gr}, 
using the Les Houches Event File (LHEF) format
\cite{Boos:2001cv,Alwall:2006yp}. If the events contain new particles, 
so-called \ttt{QNUMBERS} blocks 
\cite{Pukhov:2005je,Allanach:2006fy,Alwall:2007mw}, described in
sec.~\ref{sec:info} below, can be used 
to add information on the quantum numbers of new particles, and
SLHA decay tables \cite{Skands:2003cj} may also be provided. 
(For SUSY models, in addition, complete SLHA spectra can be given, as
discussed in sec.~\ref{sec:susy}.) The encoding of colour flow is then
particularly important, for the events to be showered and hadronized
correctly. Some pedagogical examples, with illustrations, are given in 
the original LHA paper~\cite{Boos:2001cv}, and further explicit
examples with colour-epsilon and colour-sextet structures are given in 
\secRef{sec:showers} below.
The LHEF paper~\cite{Alwall:2006yp} describes how to encode this in an 
LHE file, with examples of correct LHE files available, e.g., in 
\Py's \texttt{examples/} directory.

When reading events from LHE files, the BSM/SLHA information may
either be enclosed within the LHE file (preferred), or provided in a
separate file. In the former case, the BSM/SLHA information should be
included in the \ttt{<header>} part of the LHE file
\cite{Alwall:2006yp}, inside an \ttt{<slha>} tag
\cite{Alwall:2007mw}. In the latter case, a separate BSM/SLHA file may
be specified using the \Py~8 command \ttt{SLHA:file = fileName}. The
mode \ttt{SLHA:readFrom} gives the user some additional control over
whether and from where BSM/SLHA information is read in. It should
normally be left at its default setting, but can optionally be used
either to switch off SLHA read-in entirely, or to force read-in from a
specific file:

{\small 
\begin{verbatim}
 SLHA:readFrom = 0 # do not read BSM/SLHA information at all
               = 1 # (default) read in from the <slha>...</slha> block of a LHEF, 
                   # if such a file is read during initialization, and else from  
                   # SLHA:file 
               = 2 # read in from SLHA:file
\end{verbatim}
}

The framework described in section~\ref{sec:slhaptr} represents a
third option which combines features from both of the two
possibilities above. It allows parameters and matrix-element code to
be imported directly from external packages, to generate semi-internal
processes in \Py~8 (i.e., without an intermediate LHE file) in a fully
automated and generic way, as long as the final-state parton
multiplicity does not exceed the limitations of \Py's internal
hard-process phase-space generator \cite{Sjostrand:2007gs}.  A working
interface between \Py~8 and \Mg~5~\cite{Alwall:2011uj} has been
constructed along these lines, for $2\to 2$ processes, and will be
reported on separately. Here, we focus on the \Py~8 side of the
interface.

The interface basically consists of two parts: 1) information about
particles and couplings using a generalized SLHA format
(section~\ref{sec:info}), and 2) accessing that information from
within a semi-internal \Py~8 process (section~\ref{sec:slhaptr}).

\subsection{Information about New Particles \label{sec:info}}

Information about particle quantum numbers, masses, couplings, and
decays, can be given in an ASCII file, using a generalization of the
SLHA \cite{Skands:2003cj} and BSM-LHEF \cite{Alwall:2007mw} formats,
whose name is provided to \Py\ by setting the word \ttt{SLHA:file =
  fileName}.

\subsubsection{QNUMBERS}
The SLHA file should contain a \ttt{QNUMBERS} block
\cite{Alwall:2007mw} for each state not already associated with an ID
code (a.k.a.~PDG code, see \cite{Sjostrand:2007gs,pdg2012}
for a list) in \Py~8. For a hypothetical electrically neutral
colour-octet self-conjugate fermion (a.k.a.~a gluino) that we wish to
assign the code 7654321 and the name ``balleron'', the structure of
this block should be

{\small
\begin{verbatim}
BLOCK QNUMBERS 7654321 # balleron
      1     0  # 3 times electric charge
      2     2  # number of spin states (2S+1)
      3     8  # colour rep (1: singlet, 3: triplet, 6: sextet, 8: octet)
      4     0  # Particle/Antiparticle distinction (0=own anti)
\end{verbatim}
}

For a non-selfconjugate particle, separate names can be given for the
particle and its antiparticle. For a heavy up-type quark,  

{\small
\begin{verbatim}
BLOCK QNUMBERS 8765432 # yup yupbar
      1     2  # 3 times electric charge
      2     2  # number of spin states (2S+1)
      3     3  # colour rep (1: singlet, 3: triplet, 6: sextet, 8: octet)
      4     1  # Particle/Antiparticle distinction (0=own anti)
\end{verbatim}
}
Note that the name(s) given after the \texttt{\#} mark in the block
definition are optional and entirely up to the user. If present, they
will be used, e.g., when printing out event records with \Py's
\texttt{event.list()} method. 

The SM quantum numbers given in the \texttt{QNUMBERS} blocks are
required by \Py~8 for QED and QCD showering, and for colour-flow
tracing. (Currently, \Py\ does not make use of the spin information.)
As a rule, we advise to avoid clashes with existing ID codes, to the
extent possible in the implementation. A useful rule of thumb is to
only assign codes higher than 3 million to new states, though one
should be careful not to choose numbers larger than a 32-bit computer
integer can contain, which puts a cap at $\sim$ 2 billion.

\subsubsection{MASS}
The file should also contain the SLHA block \ttt{MASS}, which must, as
a minimum, contain one entry for each new state, in the form

{\small
\begin{verbatim}
BLOCK MASS
#     ID code  pole mass in GeV 
      7654321   800.0  # m(balleron)
      8765432   600.0  # m(yup)
\end{verbatim}
}

In principle, the block can also contain entries for SM
particles. Here, some caution and common sense must be applied,
however. Allowing SLHA spectra to change hadron and/or light-quark
masses in \Py~8 is strongly discouraged, as these parameters are used
by the parton-shower and hadronization models. Changing the $b$-quark
mass, for instance, should ideally be accompanied by a retuning of the
$b$ fragmentation parameters. Since this is not the sort of question a
BSM phenomenology study would normally address, by default, therefore,
\Py~8 tries to protect against unintentional overwriting of the SM
sector via the flag \ttt{SLHA:keepSM}, which is \ttt{on} by
default. To be more specific, this flag causes particle data
(including decay tables, see below) for ID codes in the ranges 1 -- 24
and 81 -- 999,999 to be ignored. Notably this includes $Z^0$ (ID 23),
$W^{\pm}$ (ID 24), and $t$ (ID 6). The SM Higgs (ID 25), however, may
still be modified by the SLHA input, as may other particles with ID
codes in the range 25 -- 80 and beyond 1,000,000.  If you switch off
this flag then also SM particles are modified by SLHA input.
  
Alternatively, the parameter \ttt{SLHA:minMassSM}, with default value
100.0 GeV, can be specified to allow any particle with ID code below
1,000,000 to be modified, if its default mass in \Py lies below some
threshold value, given by this parameter. The default value of 100.0
allows SLHA input to modify the top quark, but not, e.g., the $Z^0$
and $W^\pm$ bosons.

\subsubsection{DECAY}
The file may also include one or more SLHA decay tables
\cite{Skands:2003cj}. New BSM particles without decay tables will be
treated as stable by \Py~8. For coloured states, this may result in
errors at the hadronization stage, and/or in the possibly
unintentional production of so-called $R$-hadrons
\cite{Fairbairn:2006gg}, with a reasonably generic model for the
latter available in \Py~8~\cite{manPage:RH}.  On the other hand, a
redefinition of \Py's treatment of the decays of SM particles, like
$Z^0$ and $W^{\pm}$ may be undesirable, since \Py's internal treatment
is normally more sophisticated (discussed briefly in
sec.~\ref{sec:sparticle_decays}). Thus, again, caution and common
sense is advised when processing (B)SM particles through \Py, with the
protection parameters \ttt{SLHA:keepSM} and \ttt{SLHA:minMassSM}
described above also active for decay tables. An option for overriding
the automatic read-in of decay tables is also provided, by setting the
flag \ttt{SLHA:useDecayTable = false}.

The format for decay tables is \cite{Skands:2003cj}

{\small
\begin{verbatim}
#          ID     WIDTH in GeV     
DECAY   7654321  2.034369169E+00  # balleron decays
#    BR                NDA      ID1      ID2      ID3
     9.900000000E-01    3        6          5        3   # BR( -> t b s )
     1.000000000E-02    3        4          5        3   # BR( -> c b s ) 
\end{verbatim}
}

Note that the branching ratios (BRs) must sum up to unity, hence
zeroing individual BRs is not a good way of switching modes
off. Instead, \Py~8 is equipped to interpret a negative BR as a mode
which is desired switched off for the present run, but which should be
treated as having the corresponding positive BR for purposes of
normalization.

Finally, a note of warning on double counting. This may occur if a
particle can decay via an intermediate on-shell resonance. An example
is $H^0\to q_1\bar{q}_2q_3\bar{q}_4$ which may proceed via $H^0\to WW$
followed by $W\to q\bar{q}'$. If branching ratios for both $H^0\to WW$
and $H^0\to q_1\bar{q}_2q_3\bar{q}_4$ are included, each with their
full partial width, a double counting of the on-shell $H^0\to WW$
contribution would result. (This would also show up as branching
ratios summing to a value greater than unity.) Such cases should be
dealt with consistently, e.g., by subtracting off the on-shell
contributions from the $H^0\to q_1\bar{q}_2q_3\bar{q}_4$ partial
width.

\subsection{Accessing the Information from a Semi-Internal Process \label{sec:slhaptr}}

Already the original SLHA1 \cite{Skands:2003cj} allowed for the
possibility to create user-defined blocks, beyond those defined by the
accord itself. The only requirement is obviously that the block names
already defined in the accord(s) should not be usurped.  The SLHA
interface in \Py~8 will store the contents of all blocks, both
standard and user-defined ones, as internal vectors of strings.

By default, \Py's internal BSM implementation only extracts numerical
content from those blocks it recognizes (i.e., the standard SLHA 1\&2
blocks and \ttt{QNUMBERS}), and uses those to initialize its couplings
and particle data arrays. However, generic methods are also provided,
that can be used access to the contents of \emph{any} block, whether
standard or user-defined, from inside any class inheriting from \Py's
\ttt{SigmaProcess} class (i.e., in particular, from any semi-internal
process written by a user), through its SLHA pointer, \ttt{slhaPtr},
by using the following methods: 

{\small 
\begin{verbatim}
  bool slhaPtr->getEntry(string blockName, double& val);  
  bool slhaPtr->getEntry(string blockName, int indx, double& val);  
  bool slhaPtr->getEntry(string blockName, int indx, int jndx, double& val);  
  bool slhaPtr->getEntry(string blockName, int indx, int jndx, int kndx, double& val); 
\end{verbatim}
} 

This particular example assumes that the user wants to read the
entries (without index, indexed, matrix-indexed, or 3-tensor-indexed,
respectively) in the user-defined block \texttt{blockName}, and that
the entry value, \ttt{val}, should be interpreted as a
\texttt{double}. In fact, the last argument is templated, and hence if
anything other than a \ttt{double} is desired to be read, the user has
only to give the last argument a different type. Since the user
presumably knows what type of content his/her own user-defined blocks
contain, this solution allows the content to be accessed in the
correct format, without \Py\ needing to know what that format is
beforehand. If anything goes wrong (i.e., the block does not exist, or
it does not have an entry with that index, or that entry cannot be
read as a \ttt{double}), the method returns \ttt{false}; \ttt{true}
otherwise. This effectively allows input of completely arbitrary
parameters using the SLHA machinery, with the user having full control
over names and conventions. Of course, it is then also the user's
responsibility to ensure complete consistency between the names and
conventions used in the SLHA input, and those assumed in any
user-written semi-internal process code.

Note also that the special SLHA block \texttt{SMINPUTS} (containing SM
parameters \cite{Skands:2003cj}) will always be accessible through the
methods above, regardless of whether a corresponding SLHA block has
been read in or not. The \texttt{SMINPUTS} block is initialized
starting from PYTHIA's own internal default values, with subsequent
modifications as dictated by updates to PYTHIA's particle and
parameter databases before initialization and/or by SLHA read-in. This
functionality is intended to give a generic BSM implementation access
to the SM parameters contained in \texttt{SMINPUTS} in a universal
way.

To give a specific example, the interface to \textsc{Madgraph}~5 was
structured in the following way. Among the possible output formats
available for matrix elements in \textsc{Madgraph}~5, one is a mode
called \texttt{pythia8}. When invoked, this mode writes out the
corresponding matrix element(s) in exactly the format required by
\Py~8's semi-internal process machinery. The resulting code can
therefore be imported directly into \Py~8, and \textsc{Madgraph} even
provides explicit instructions and a Makefile for doing precisely
that.  In general, however, such matrix elements may contain
parameters that refer, e.g., to couplings in a model unknown to \Py. A
central question was therefore how to provide information on such
parameters at runtime, in a sufficiently generic manner. The solution
is that \textsc{Madgraph} writes out the relevant parameters as
custom-made SLHA-like blocks in a BSM/SLHA file included together with
the matrix-element code. It then also inserts appropriate calls to
\texttt{slhaPtr->getEntry()} in the cross-section expressions, so that
each parameter can be retrieved when needed, without any user
intervention required.

Note that this entirely circumvents a particularly troublesome issue
that before was thought to make any truly universal ``BSM Accord''
impractical, the problem of agreeing on a common standard for names
and parameters for completely arbitrary models. In the
\textsc{Madgraph}-\textsc{Pythia} interface, it is sufficient that
\textsc{Madgraph} itself assigns some unique names and contents to
each block. It has complete freedom in choosing which conventions to
use, as long as it correctly inserts the corresponding
\ttt{readEntry()} calls in its matrix-element output. This effectively
generalizes the SLHA data structure to apply to completely general BSM
models.

The interface has been tested by the authors (in collaboration with
our \textsc{Madgraph} colleagues) to work for importing both a few
trivial examples of models, such as a 4-generation model, to more
exotic ones, such as a model with colour-sextet diquarks and one with
a baryon-number-violating vertex. Showers in such topologies are the
topic of section~\ref{sec:showers}.

\section{Showers and Hadronization in Exotic Colour Topologies\label{sec:showers}}

In this section, we describe \Py's treatment of QCD radiation in
topologies containing colour-epsilon tensors (\secRef{sec:epsilon})
and ones involving particles with colour-sextet quantum numbers
(\secRef{sec:sextets}). This applies regardless of whether the event
is generated as an internal, semi-internal, or LHEF process. We also
comment briefly on hadronization aspects, pointing out relevant
sources of further information.

\begin{figure}[tp]
\centering
\subfloat{
\small
\begin{tabular}[t]{ccrrrrr}
\multicolumn{7}{c}{\includegraphics*[scale=0.65]{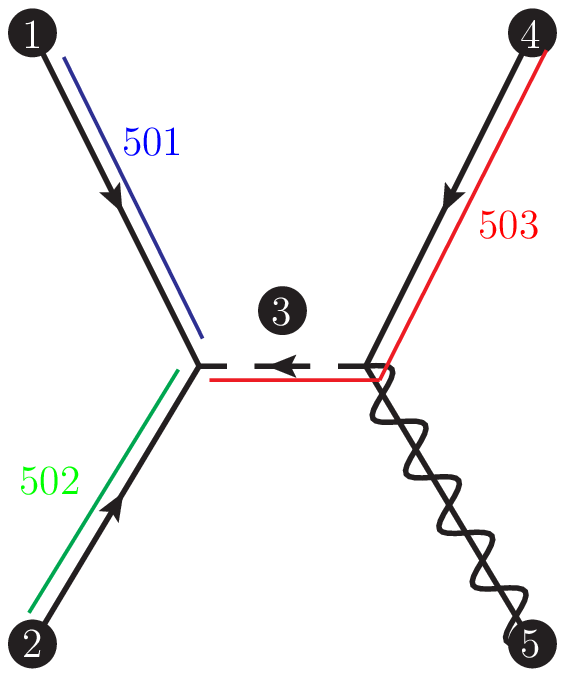}}\\
I & IST &  {\small ID} & \multicolumn{2}{c}{Parents}  & \multicolumn{2}{c}{Colours} \\
\includegraphics*[scale=0.62]{LH1} &
-1 & 3 & 0 & 0 & 501 & 0\\
\includegraphics*[scale=0.62]{LH2}
 & -1 & 5 & 0 & 0 & 502 & 0\\
\includegraphics*[scale=0.62]{LH3}
 & 2 &-1000006 & 1 & 2 & 0 & 503\\
\includegraphics*[scale=0.62]{LH4}
 & 1 & -6 & 3 & 3 & 0 & 503\\
\includegraphics*[scale=0.62]{LH5}
 & 1 & 1000022 & 3 & 3 & 0 & 0\\
\end{tabular}}
\hspace*{5mm}
\subfloat{
\small \begin{tabular}[t]{ccrrrrr}
\multicolumn{7}{c}{\includegraphics*[scale=0.65]{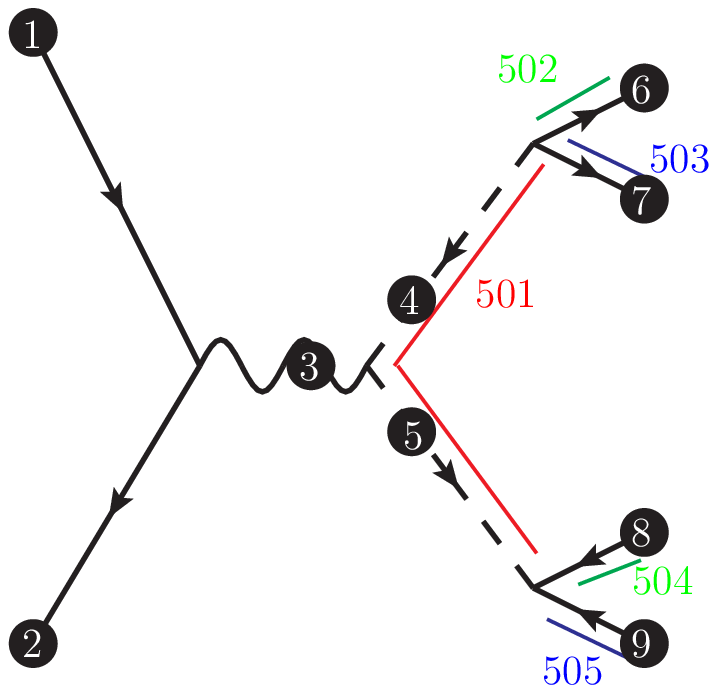}}\\
I & IST &  {\small ID} & \multicolumn{2}{c}{Parents}  & \multicolumn{2}{c}{Colours} \\
\includegraphics*[scale=0.62]{LH1} &
-1 & 11 & 0 & 0 & 0 & 0\\
\includegraphics*[scale=0.62]{LH2} &
-1 & -11 & 0 & 0 & 0 & 0\\
\includegraphics*[scale=0.62]{LH3} &
2 & 23 & 1 & 2 & 0 & 0\\
\includegraphics*[scale=0.62]{LH4} &
2 & -1000006 & 3 & 3 & 0 & 501\\
\includegraphics*[scale=0.62]{LH5} &
2 & 1000006 & 3 & 3 & 501 & 0\\
\includegraphics*[scale=0.62]{LH6} &
1 & 3 & 4 & 4 & 502 & 0\\
\includegraphics*[scale=0.62]{LH7} &
1 & 5 & 4 & 4 & 503 & 0\\
\includegraphics*[scale=0.62]{LH8} &
1 & -3 & 5 & 5 & 0 & 504\\
\includegraphics*[scale=0.62]{LH9} &
1 & -5 & 5 & 5 & 0 & 505
\end{tabular}
}\\
\subfloat{
\small \begin{tabular}[t]{ccrrrrr}
\multicolumn{7}{c}{\includegraphics*[scale=0.65]{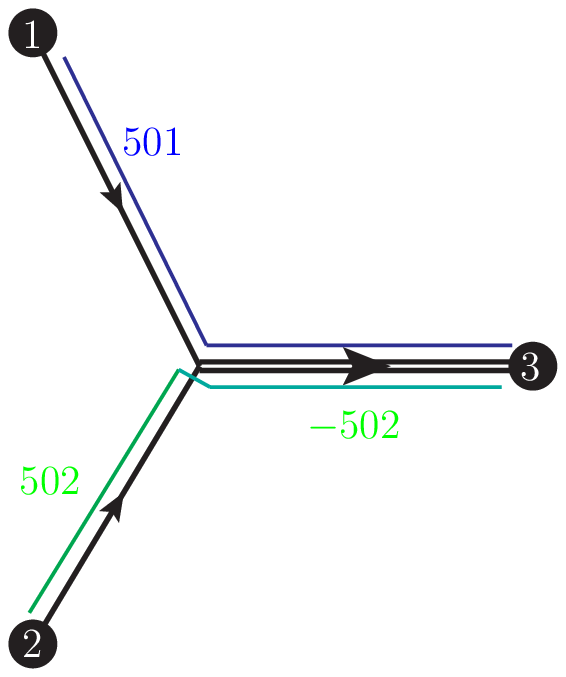}}\\
I & IST &  {\small ID} & \multicolumn{2}{c}{Parents}  & \multicolumn{2}{c}{Colours} \\
\includegraphics*[scale=0.62]{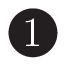} &
-1 & 1 & 0 & 0 & 501 & 0\\
\includegraphics*[scale=0.62]{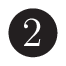} &
-1 & 1 & 0 & 0 & 502 & 0\\
\includegraphics*[scale=0.62]{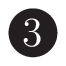} &
1 & 6000001 & 1 & 2 & 501 & -502
\end{tabular}
\label{fig:sextet}
}
\hspace*{5mm}
\subfloat{\small 
\begin{tabular}[t]{ccrrrrr}
\multicolumn{7}{c}{\includegraphics*[scale=0.65]{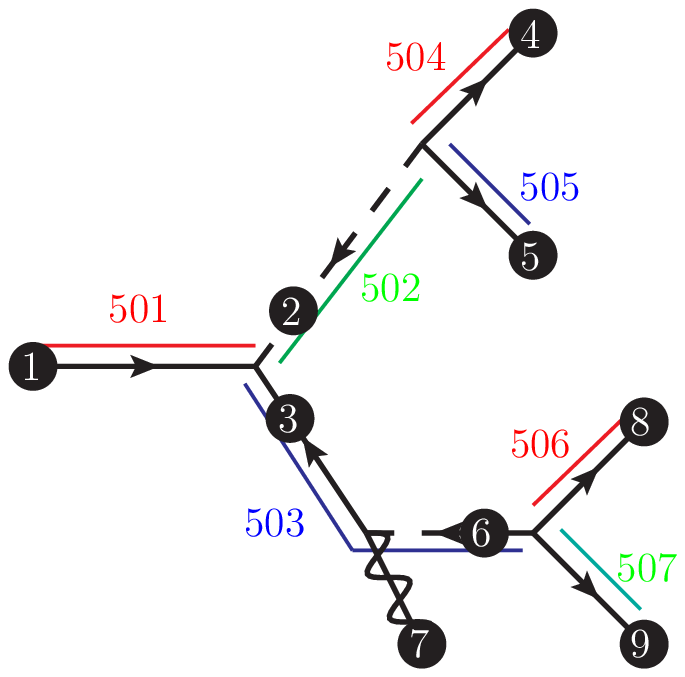}}\\
I & IST &  {\small ID} & \multicolumn{2}{c}{Parents}  & \multicolumn{2}{c}{Colours} \\
\includegraphics*[scale=0.62]{LH1.eps} &
2 & 8 & 0 & 0 & 501 & 0\\
\includegraphics*[scale=0.62]{LH2.eps} &
2 &-1000006 & 1 & 1 & 0 & 502\\
\includegraphics*[scale=0.62]{LH3.eps} &
2 & -7 & 1 & 1 & 0 & 503\\
\includegraphics*[scale=0.62]{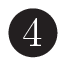} &
1 & 5 & 2 & 2 & 504 & 0\\
\includegraphics*[scale=0.62]{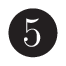} &
1 & 3 & 2 & 2 & 505 & 0\\
\includegraphics*[scale=0.62]{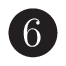} &
2 & -1000006 & 3 & 3 & 0 & 503\\
\includegraphics*[scale=0.62]{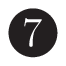} &
1 & 1000024 & 3 & 3 & 0 & 0\\
\includegraphics*[scale=0.62]{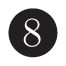} &
1 & 5 & 6 & 6 & 506 & 0\\
\includegraphics*[scale=0.62]{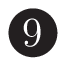} &
1 & 3 & 6 & 6 & 507 & 0
\end{tabular}
\label{fig:triple}
}
\caption{Illustration of the assignment of Les Houches colour tags in
  four different exotic colour topologies. Lines
  corresponding to colour (anticolour) tags are drawn above (below) the
  propagators. {\sl Top Left:} $sb \to \tilde{t}^*
  \to \bar{t} \tilde{\chi}_1^0$. {\sl Top Right:} $e^-e^+\to Z^0 \to
  (\tilde{t}^*\to sb)(\tilde{t}\to\bar{s}\bar{b})$. {\sl Bottom Left:}  
  production of a colour-sextet particle,  assigned the fictitious
  ID code 6000001; the negative anti-colour tag (drawn below the
  sextet propagator) is interpreted as an additional (positive) colour
  tag. {\sl Bottom Right:} A complicated
  baryon-number-violating  cascade decay (of a hypothetical
  fourth-generation fermion) producing a total of three
  colour-connected baryon-number-violating vertices; such topologies
  (with three or more interconnected colour junctions) cannot yet be
  handled by \Py's string fragmentation model \cite{Sjostrand:2002ip}.
\label{fig:exotic}}
\end{figure}

As an aid to implementations using LHEF, a few examples of how to
arrange Les Houches colour tags in colour-epsilon and colour-sextet
cases are given in \FigsRef{fig:exotic} and \ref{fig:sextet},
respectively.  For completeness, we also show the status and ID codes,
and the mother information, for each particle. These are explained in
more detail in \cite{Boos:2001cv}, where also more illustrations
(including both standard and baryon-number violating ones) can be
found.

For completeness, \FigRef{fig:triple} shows a situation which \Py is
not yet capable of handling. The illustration shows a complicated
baryon-number-violating cascade decay of a hypothetical
fourth-generation top quark (assigned ID code 8) involving both
supersymmetric, fourth-generation, and SM particles, to produce a
situation with a total of three colour-connected
baryon-number-violating vertices. At the moment, \Py's junction
fragmentation model \cite{Sjostrand:2002ip} is at most capable of
handling up to two connected colour junctions (specifically, single
junctions and junction-antijunction systems), hence only if a $g\to
q\bar{q}$ branching in the shower happens to break up the
triple-junction system into smaller colour-singlet subsystems would
\Py's fragmentation model be able to deal with it.

A somewhat less pathological case in which multi-junction topologies
may result is if a single baryon-number violating vertex becomes
colour-connected to both of the junctions in the (baryon) beam
remnants. This may happen some small fraction of the time through
multiple parton interactions.  In such cases, the following error
message will be printed and the generation of the event restarted,

{\noindent\footnotesize
\begin{verbatim}
  Error in ColConfig::insert: junction topology too complicated; too many junction legs   
\end{verbatim}
}

\subsection{Colour-Epsilon Topologies \label{sec:epsilon}}

For colour topologies involving the epsilon tensor in colour space
(i.e., colour topologies with non-zero baryon number) we first
consider the example of $\tilde{t}\to \bar{q}\bar{q}$ in the RPV-SUSY
model.

The Lagrangian for the UDD-type interaction terms is
\begin{equation}
\mathcal L = -\lambda''_{ijk} \epsilon^{lmn} \left(\tilde u^l_{Ri}
(\bar
d^c)^m_j P_R d^n_k + \tilde d^m_{Rj} (\bar u^c)^l_i P_R d^n_k + \tilde
d^n_{Rk} (\bar u^c)^l_i P_R d^n_k + h.c. \right)
\end{equation}

To extract the behaviour of the radiation function, we look at the
ratio of exact matrix element for $\tilde t_R(p_1) \rightarrow \bar
d(p_2) \bar s(p_3) + g(q)$ via $\lambda''_{312}$ to the matrix element
for $\tilde t_R(p_1) \rightarrow \bar d(\hat{p}_2) \bar s(\hat{p}_3)$ and
retaining only the parts that are soft- or collinear-singular (i.e.,
which diverge for one or more $q\cdot p_i\to 0$). 
Since momentum  is explicitly conserved in the
shower branching process, the pre- and post-emission momenta must be related by 
\begin{equation}
p_1 = \hat{p}_2 + \hat{p}_3 = p_2 + p_3 + q~,
\end{equation}
with $p_1^2 = m_1^2 = \hat{s}$ the invariant mass of the decaying squark.

The Born-level matrix element
squared is given by:
\begin{equation}
|M_{0}|^2 \ = \ |\lambda''_{312}|^2 \, (N_c-1)!\, \hat{s}
\end{equation}

\begin{figure}[tp]
\centering
\includegraphics[width=40mm]{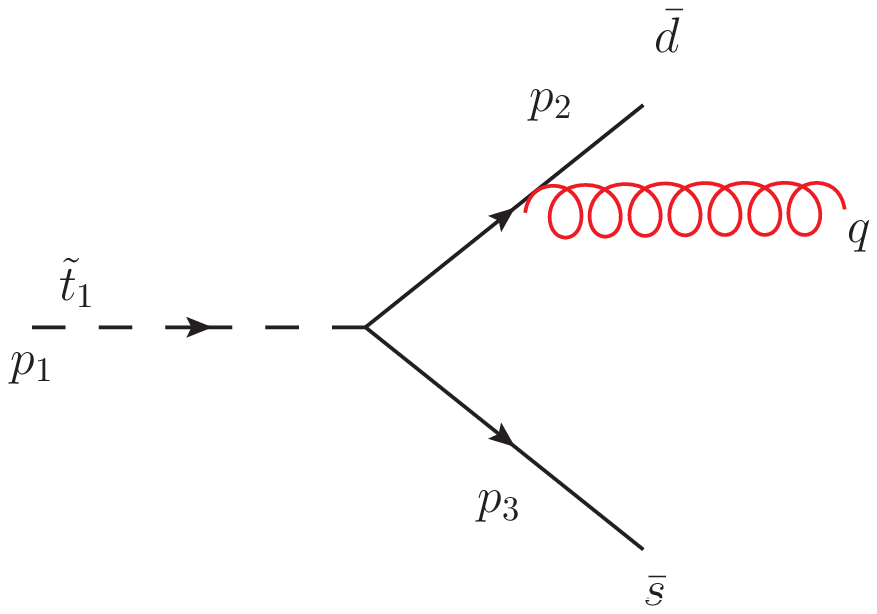}
\includegraphics[width=40mm]{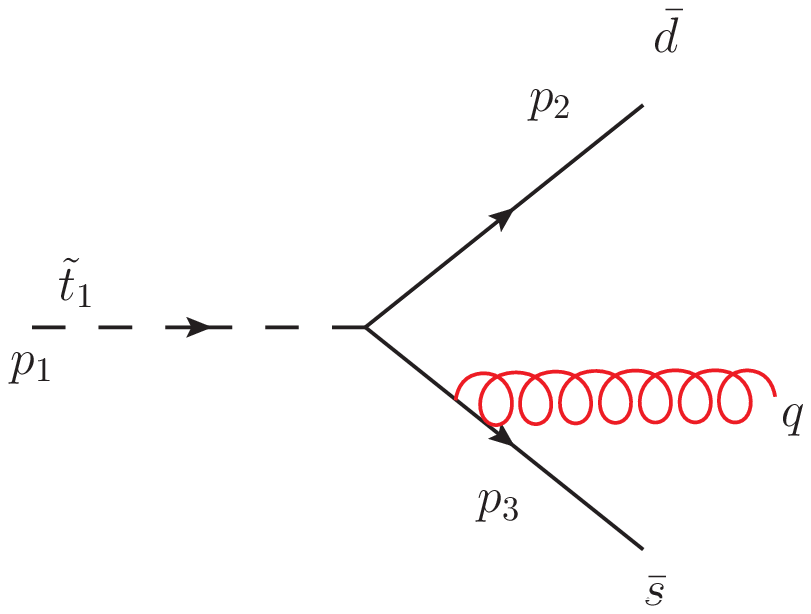}
\includegraphics[width=40mm]{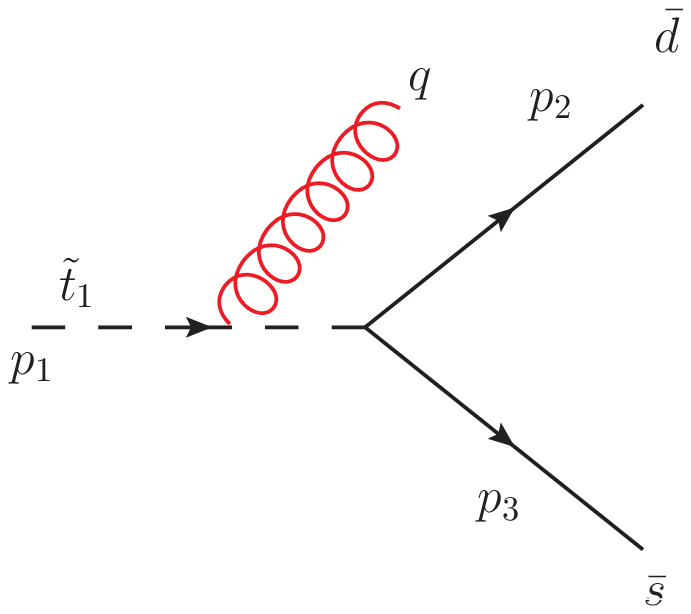}
\caption{\label{fig:feyn} Gluon emission from RPV vertices with
  $\epsilon$-tensor.}
\end{figure}

Three diagrams (shown in Fig \ref{fig:feyn}) contribute to the process
where one gluon is emitted from this configuration.  The matrix
element corresponding to this process i.e.\ $\tilde t_R(p_1)
\rightarrow \bar d(p_2) \bar s(p_3) g(q)$ is denoted by $M_1$ and, for
massless decay products ($p_2^2 = p_3^2 =
0$), is given by
\begin{eqnarray}
\nonumber |M_1|^2 & =   & 2g_s^2|\lambda''|^2 (N_C-1)! C_F \hat s \Bigg[ \\
\nonumber & & \quad \frac{1}{N_C-1}\left( \frac{(p_1 \cdot p_2) }{(p_1 \cdot q)(p_2 \cdot q)} + \frac{(p_1 \cdot p_3) }{(p_1 \cdot q)(p_3 \cdot q)}+ \frac{(p_2 \cdot p_3)}{(p_2 \cdot q)(p_3 \cdot q)} \right) \\
&& \quad + \frac{1}{\hat{s}}\left(\frac{(p_2 \cdot q)}{(p_3 \cdot q)}
+   \frac{(p_3 \cdot q)}{(p_2 \cdot q)} + \frac{X}{N_c -1}
+ Y \right) \Bigg] ~,\label{eq:me}
\end{eqnarray}
where 
\begin{eqnarray}
\nonumber X & = & 10 - \frac{6m_1^2}{p_1 \cdot q}  
- (p_2 \cdot p_3) \left\{  \frac{(p_2 \cdot q)}{(p_1 \cdot q)(p_3  \cdot q)} + \frac{(p_3 \cdot q)}{(p_1 \cdot q)(p_2 \cdot q)} + \frac{(p_1 \cdot q)}{(p_2 \cdot q)(p_3 \cdot q)} \right\}~, \\
Y & = & -\frac{(p_2 \cdot p_3)(m^2 - p_1 \cdot q)}{(p_1 \cdot q)^2}~.
\end{eqnarray}
The antenna pattern represented by \eqRef{eq:me} can be characterized
as follows: the terms on the second line represent three
soft-eikonal dipole factors (see, e.g., \cite{Buckley:2011ms}), 
one for each of the three possible two-particle combinations. The
factor $1/(N_c-1)$ in front of the dipole factors  
implies that 
the normalization of each of these eikonals is half as large as that
of the eikonal term in an ordinary $q\bar{q}$ antenna, see, e.g.,
\cite{Gustafson:1987rq,GehrmannDeRidder:2005cm,Ridder:2011dm}. 
The two first terms on the last line of \eqRef{eq:me} 
correspond to additional purely collinear singularities for each of
the quarks. The factor $1/(N_c-1)$ is here absent; the collinear
singularities have the same strength as those of an ordinary
$q\bar{q}$ antenna. The $X$ and $Y$ terms contain subleading-color and
a quasi-collinear term for the decaying $\tilde{t}$, respectively.

The eikonal terms (including the leading part of the $Y$ term,
$\propto m^2/(p_1\cdot q)^2$) agree with the expression in
\cite{Dreiner:1999qz,Richardson:2000nt}, which is used  
 to generate radiation for this type
of colour topology in \Hw~\cite{Corcella:2000bw}. 
(Note that, in the  \Hw\ implementation, the pattern is 
generated using ordinary full-strength radiation functions, by
selecting randomly between each two-particle combination, thereby
reproducing the full pattern when summing over
events~\cite{Dreiner:1999qz,Richardson:2000nt}.)  

For the implementation in \Py, we have chosen a different strategy, as
follows. 
First, using momentum conservation, we may rewrite the antenna 
pattern above to only contain the final-state particle momenta,
\begin{equation}
\frac{p_1 \cdot p_2}{(p_1 \cdot q)(p_2 \cdot q)} + \frac{p_1 \cdot p_3}{(p_1 \cdot q)(p_3 \cdot q)} = 
\frac{p_2 \cdot p_3}{(p_2 \cdot q)(p_3 \cdot q)} + \frac{2}{p_1 \cdot q}
\end{equation}
This reduces the eikonal part of expression to a single antenna between the
two final-state quarks, plus subleading leftover terms. 
The eikonal and the collinear terms then correspond exactly to the
standard radiation pattern from a $q \bar q$ dipole with an extra term
of $\mathcal{O}(\frac{1}{N_c})$.  

For the present work, we therefore
take the radiation pattern of a standard-strength
dipole spanned between the two final-state 
quarks as our starting point. Using $s_{ij} = 2p_i \cdot p_j$, this
radiation function is given by~\cite{Gustafson:1987rq}
\begin{eqnarray}
\frac{|M_{Z\to q\bar{q}+g}|^2}{|M_{Z\to q\bar{q}}|^2} & = & 8 \pi \alpha_s C_F \left(\frac{2s_{23}}{s_{2q}s_{3q}} +
\frac{s_{2q}}{\hat{s} s_{3q}} + \frac{s_{3q}}{\hat{s} s_{2q}} \right)~.
\label{eq:leadingN}
\end{eqnarray}
The \Py\ showers are not based directly on
\eqRef{eq:leadingN}, but rather on Altarelli-Parisi (AP) 
splitting kernels, which partition the radiation pattern onto two
terms, each of which is governed by the $q\to qg$ splitting function, 
\begin{equation}
P_{q\to qg}(z) = C_F \frac{1+z^2}{1-z}~,
\end{equation}
with $z$ the energy fraction retained by the quark after emitting the
gluon. The energy fractions of the final-state quarks, $2$ and $3$, 
are defined as in \cite{Bengtsson:1986et},
\begin{equation}
z_{2}~=~\frac{x_2}{x_2+x_q}~=~\frac{m_1^2-s_{3q}}{m_1^2+s_{2q}}
~~~~~;~~~~~
z_{3}~=~\frac{x_3}{x_3+x_q}~=~\frac{m_1^2-s_{2q}}{m_1^2+s_{3q}}~.
\end{equation}
The expression actually used in the \Py\ showering is the sum of the 
AP contributions, 
\begin{equation}
\frac{|M_{1}|^2}{|M_{0}|^2} ~\stackrel{\Py}{\sim}~ 
8\pi\alpha_s
\left( 
   \frac{P(z_2)}{s_{2q}}
 + \frac{P(z_3)}{s_{3q}}
\right)~.\label{eq:pythiaPattern}
\end{equation}

\begin{figure}[t!]
\centering
\begin{tabular}{rr}
\multicolumn{1}{c}{\centering $m = 300$ GeV, $E_g = 10$ GeV}
&
\multicolumn{1}{c}{\centering $m = 300$ GeV, $\theta_{qg} = 20^\circ$}
\\
\includegraphics*[scale=0.834]{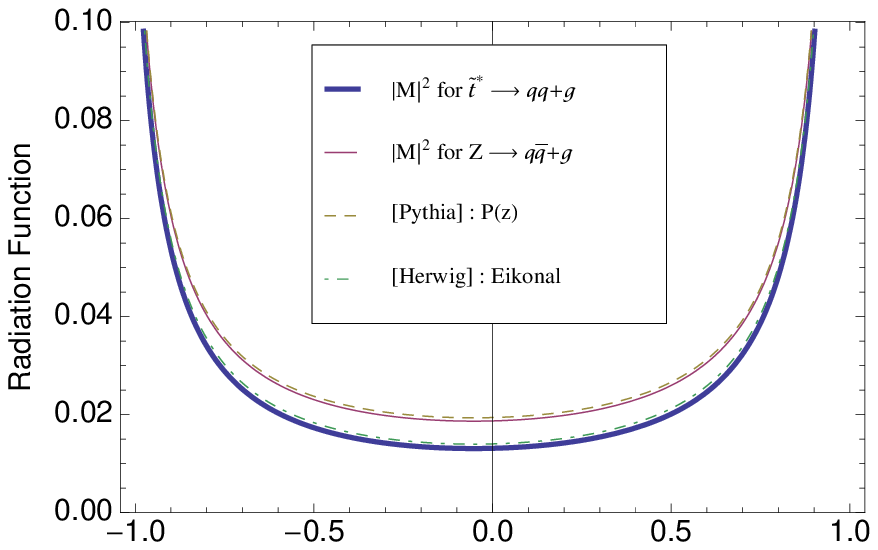} & 
\includegraphics*[scale=0.834]{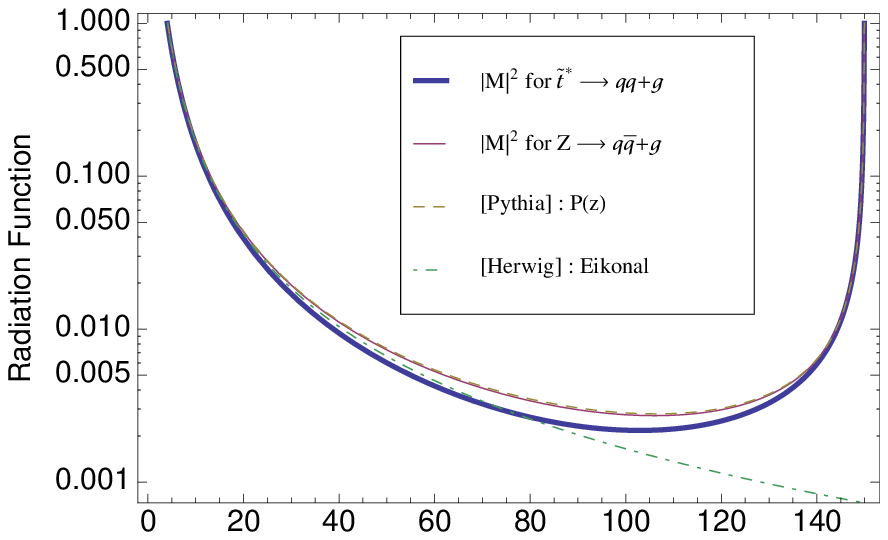}
\\
\includegraphics*[scale=0.778]{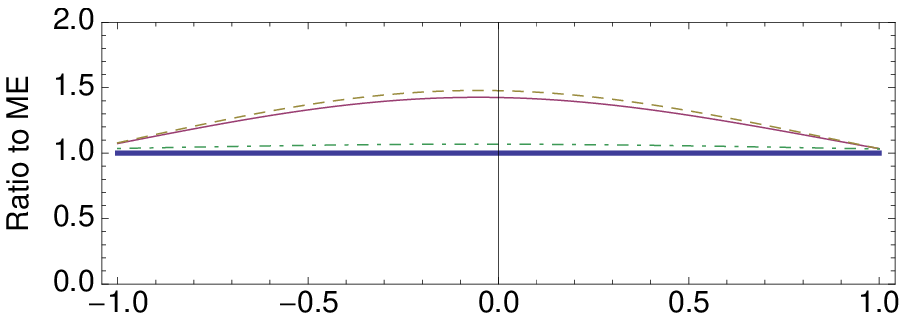} & 
\includegraphics*[scale=0.778]{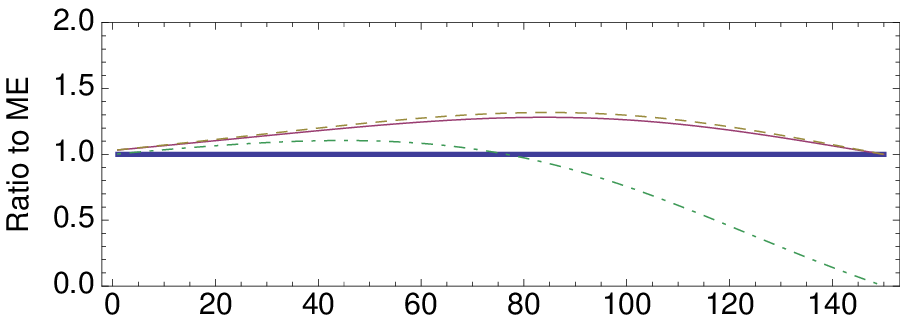}\\
\footnotesize$\cos\theta_{qg}$ & \footnotesize$E_g$ [GeV]
\end{tabular}
\caption{Illustration of radiation functions for gluon emission 
  in $\tilde{q}^*\to q_1q_2$ decays for soft (left) and collinear (right)
  gluons. 
\label{fig:radPatterns}}
\end{figure}
The full matrix-element ratio, $|M_1|^2/|M_0|^2$, as well as the
various approximate forms discussed here, are illustrated in
Fig.~\ref{fig:radPatterns}, with the mass of the decaying $\tilde{t}$
arbitrarily set to $m_1 = 300$ GeV. On the left-hand pane, we show the
size of the radiation function (without the overall factor
$8\pi\alpha_sC_F$) as a function of the opening 
angle between the final-state gluon and one of the quarks, for a
fixed ($\sim$soft) gluon energy $E_g = 10$ GeV. On the right-hand pane, we
show the dependence on energy, for a fixed ($\sim$collinear) opening angle
$\theta_{qg} = 20^\circ$. The bottom row shows the ratio of each 
approximation to the matrix-element result.

The thick solid (blue) line represents 
the full $\tilde{t}^* \to qq+g$ matrix element, \eqRef{eq:me}. For comparison, 
the thin solid (red) line shows the pattern obtained for a 
standard dipole, \eqRef{eq:leadingN}. The dashed (brown) curve shows the
\Py\ approximation to the dipole pattern, given by 
the sum of the AP splitting kernels in
\eqRef{eq:pythiaPattern}. Finally, the light dot-dashed (green) curve
shows the eikonal approximation to the matrix element, used by \Hw. 

In the soft limit (left-hand pane of \FigRef{fig:radPatterns}), all
the expressions agree in the two extremal points, in which the gluon
is \emph{both} soft and collinear. 
For wide-angle soft emissions, e.g.\ at 90$^\circ$ opening
angle, the standard dipole pattern (as well as its DGLAP variant)
overestimate the full matrix element by up to a factor $\sim
1.5$. That is, the \Py\ shower will generate slightly too many soft
wide-angle gluons. By contrast, as would be expected in the soft limit, 
the eikonal approximation works well for all opening angles. 

In the collinear limit (right-hand pane of \FigRef{fig:radPatterns}),
the $x$ axis is now the gluon energy, with the opening angle held
fixed. All the expressions again agree for small gluon energies, in
the double soft- and collinear limit. For intermediate gluon energies,
the standard dipole pattern (as well as its DGLAP variant) again
slightly overestimate the full matrix element, while they again agree
with the matrix element in the hard collinear limit, on the right-hand
edge of the plot. 
The eikonal, however, does not include the collinear-singular terms on
the last line of \eqRef{eq:me} and hence does not reproduce 
the rise of the other curves in the hard-collinear limit. 

In summary, our shower model will slightly overestimate the total
amount of radiation, in particular at large angles, 
 while the \Hw\ model underestimates
hard-collinear radiation. We therefore regard the two as 
complementary. 
We note that the neglected terms could still 
subsequently be incorporated into \Py~8 as a matrix-element correction
\cite{Bengtsson:1986hr,Norrbin:2000uu}, presumably mostly relevant if
$B$-violating processes should indeed be observed in nature.

For the case $\tilde \chi_1^0 \rightarrow q q q $, the corresponding
expression is similar to equation (\ref{eq:me}) with three half-strength
eikonals between the quark from the neutralino-quark-squark vertex and
the two quarks from the RPV vertex \cite{Dreiner:1999qz}. Besides
these, each quark has the corresponding full collinear singularity.
\Hw\ treats this situation by randomly connecting each quark in the
final state to either of the other two quarks. We have chosen instead
to implement it in \Py~8 as three genuinely half-strength dipoles
spanned between the three final-state quarks. 

For the case of three-body RPV gluino decay, $\tilde{g}\to q q q$,
only the resonant parts, $\tilde{g}\to q \tilde{q}^*
\to q q q$, have so far been implemented in \Py~8, cf.\ section
\ref{sec:sparticle_decays}. For future off-shell contributions, 
the emission structure of the non-resonant parts will be obtained from the
relative strengths of the intermediate off-shell $\tilde{g}\to q
\tilde{q}^*$ contributions.

In all cases, the subsequent hadronization phase makes use of \Py's
ability to handle string topologies including colour junctions
\cite{Sjostrand:2002ip}, and hence issues such as baryon-number flow
should be treated at least semi-realistically, allowing studies at the
individual-particle level.

\subsection{Colour-Sextet Particles \label{sec:sextets}}

Within the leading-$N_c$ dipole approach to radiation adopted in \Py,
we represent a 
colour-sextet charge as the sum of two colour-triplet
charges, in much the same way as octet charges (e.g., gluons) 
are represented as the sum of a triplet and an antitriplet charge.  
Each triplet charge is independently colour-connected to an
antitriplet charge. Hence a sextet may be colour-connected either by a
``double bond'' to an anti-sextet (in an
overall singlet $6\bar{6}$ configuration), 
or by two ``single bonds''  to two independent antitriplet
charges, depending on the colour flow in the event. Each such ``bond''
is interpreted as an ordinary QCD dipole, with the sextet end treated
as a massive quark.

At the technical level, we note that the Les Houches colour-tag 
standard was not originally 
designed to deal with sextet colour configurations. This is easy to
remedy, however. Since a sextet never carries an anticolour, its
anticolour tag is effectively available for use. To distinguish
an additional colour (i.e., a sextet) from the ordinary anticolour
(octet) case, we adopt the convention that a negative
anticolour tag is interpreted as an additional colour, and vice versa
for anti-sextets, as was illustrated in \FigRef{fig:sextet}. This
appears to violate no present use of the standard (negative colour
tags were so far never used in practise, as far as we are aware).

We note that a more complete treatment of the radiation and
phenomenology of sextet diquarks was published while this manuscript
was in preparation, see \cite{Richardson:2011df}.

\section{Summary and Conclusions \label{sec:conclusions}}
We describe the implementation of Supersymmetry in the Monte Carlo
event generator \Py~8. We use the generic super-CKM basis of \cite{Allanach:2008qq} 
which allows CP and flavour violation in the squark sector. 
We also allow R-parity violation in production processes and decays
and the extension of the MSSM to the nMSSM.  The current
implementation includes all pair-production processes with  gluinos,
squarks, neutralinos or charginos in the final state.  We also
implement the resonant production of squarks via R-parity violating
vertices that can be relevant at a hadron collider like the LHC.
Two-body decays of all SUSY particles (except the Higgs sector) and
R-parity violating decays of neutralinos via the $\lambda''$ couplings
have been implemented.  The Higgs decays will be implemented as a part
of a future update. 

We also describe the enhancements made to the SLHA interface to allow
external programs to pass non-standard information blocks to \Py~8. 
The modifications provide a mechanism for so-called semi-internal
processes in \Py~8 to access all information read in via the SLHA
interface.  This interface can therefore be used for implementation of
generic BSM models without requiring a previous agreement on
standardization of names and parameters.  

Finally, we have commented on how \Py~8 handles showering in 
non-standard colour topologies, such as the epsilon topologies
encountered in R-parity violating models and in sextet di-quark ones.

\section*{Acknowledgements}

The authors are grateful to T.~Sj\"ostrand for the development of
\Py~8 and for useful suggestions on the implementations reported on in
this paper. We also thank B.~Fuks and S.~Mrenna for help with
validating the SUSY cross sections 
against \textsc{Xsusy} and \Py~6, respectively, and
J.~Alwall, C.~Duhr, and O.~Mattelaer for development of 
the generic BSM/SLHA framework on the \Mg\ side, and
for examples of LHE files containing 
exotic colour structures.
ND would like to thank Shailesh Lal for discussions and the CERN
theory department for hospitality during part of this work.  This work
was supported in part by the Marie Curie 
research training network ``MCnet'' (contract number
MRTN-CT-2006-035606).   

\clearpage \section*{Appendix A: Test Sparticle Spectrum}

All validations have been performed using point SPS1a (mSUGRA
parameters $m_0=250$, $m_{1/2}=100$,$A_0=0$,$\mu>0$ and $\tan \beta = 10$).
However, since the masses and mixings of superparticles at low scale
depend on renormalization group running, we give here the complete
list of masses and mixing matrices used in our validations.  The following spectrum was generated using SoftSUSY 2.0.5 \cite{Allanach:2001kg}

\begin{footnotesize}
\begin{center}
\begin{tabular}{|r|r|rrrrrr|}
\hline
   & & & & & & & \\
 PDG code   &  M(GeV)  &  Mixing     &                          &                           &             &           &            \\
   & & & & & & & \\
$\tilde g$  & & & & & & & \\
  1000021   &   607.714  &              &                          &                           &             &           &            \\
   & & & & & & & \\
 $\tilde \chi^0_i $ &     &      $ \tilde B$  &              $ \tilde W_3$  &                $\tilde H_1$  &  $\tilde H_2$  &           &            \\
  1000022  &    96.688  &       0.986  &                  -0.053  &                    0.146  &     -0.053  &           &            \\
  1000023  &   181.088  &       0.099  &                   0.945  &                   -0.270  &      0.156  &           &            \\
  1000025  &  -363.756  &      -0.060  &                   0.088  &                    0.696  &      0.710  &           &            \\
  1000035  &   381.729  &      -0.117  &                   0.311  &                    0.649  &     -0.684  &           &            \\
   & & & & & & & \\
 $\tilde \chi^+_i$  &     &           U  &                          &                   &    V         &           &            \\
           &            &       $\tilde W$  &         $\tilde H$  &                     &     $\tilde W$  &   $ \tilde H$ &            \\
  1000024  &   181.696  &       0.917  &                  -0.399  &                     &    0.973   &        -0.233    &            \\
  1000037  &   379.939  &       0.399  &                   0.917  &                     &     0.233   &         0.973   &            \\
   & & & & & & & \\
  $\tilde d$  &         &      $\tilde d_L$  &      $\tilde s_L$  &             $\tilde b_L$  & $\tilde d_R$ &$\tilde s_R$ & $\tilde b_R$  \\
  1000001  &   568.441  &       1.000  &                   0.000  &                    0.000  &      0.000  &    0.000  &     0.000  \\
  1000003  &   568.441  &       0.000  &                   1.000  &                    0.000  &      0.000  &    0.000  &     0.000  \\
  1000005  &   513.065  &       0.000  &                   0.000  &                    0.939  &      0.000  &    0.000  &     0.345  \\
  2000001  &   545.228  &       0.000  &                   0.000  &                    0.000  &      1.000  &    0.000  &     0.000  \\
  2000003  &   545.228  &       0.000  &                   0.000  &                    0.000  &      0.000  &    1.000  &     0.000  \\
  2000005  &   543.727  &       0.000  &                   0.000  &                   -0.345  &      0.000  &    0.000  &     0.939  \\
   & & & & & & & \\
 $\tilde u$&            &$\tilde u_L$  &           $ \tilde c_L$  &             $\tilde t_L$  &$ \tilde u_R$  &   $\tilde c_R$  &   $ \tilde t_R$  \\
  1000002  &   561.119  &       1.000  &                   0.000  &                    0.000  &      0.000  &    0.000  &     0.000  \\
  1000004  &   561.119  &       0.000  &                   1.000  &                    0.000  &      0.000  &    0.000  &     0.000  \\
  1000006  &   399.668  &       0.000  &                   0.000  &                    0.554  &      0.000  &    0.000  &     0.833  \\
  2000002  &   549.259  &       0.000  &                   0.000  &                    0.000  &      1.000  &    0.000  &     0.000  \\
  2000004  &   549.259  &       0.000  &                   0.000  &                    0.000  &      0.000  &    1.000  &     0.000  \\
  2000006  &   585.786  &       0.000  &                   0.000  &                    0.833  &      0.000  &    0.000  &    -0.554  \\
   & & & & & & & \\
$ \tilde e$ &           &$\tilde e_L$  &          $\tilde \mu_L$  &          $\tilde \tau_L$  & $\tilde e_R$  & $ \tilde \mu_R$  &  $\tilde \tau_R$  \\
  1000011  &   202.916  &       1.000  &                   0.000  &                    0.000  &      0.000  &    0.000  &     0.000  \\
  1000013  &   202.916  &       0.000  &                   1.000  &                    0.000  &      0.000  &    0.000  &     0.000  \\
  1000015  &   134.491  &       0.000  &                   0.000  &                    0.282  &      0.000  &    0.000  &     0.959  \\
  2000011  &   144.103  &       0.000  &                   0.000  &                    0.000  &      1.000  &    0.000  &     0.000  \\
  2000013  &   144.103  &       0.000  &                   0.000  &                    0.000  &      0.000  &    1.000  &     0.000  \\
  2000015  &   206.868  &       0.000  &                   0.000  &                    0.959  &      0.000  &    0.000  &    -0.282  \\
   & & & & & & & \\
   $\tilde \nu$  &            &  $\tilde \nu_e$  &  $\tilde \nu_{\mu}$  &  $\tilde \nu_{\tau}$  &             &           &            \\
  1000012  &   185.258  &       1.000  &                   0.000  &                    0.000  &             &           &            \\
  1000014  &   185.258  &       0.000  &                   1.000  &                    0.000  &             &           &            \\
  1000016  &   184.708  &       0.000  &                   0.000  &                    1.000  &             &           &            \\
\hline
\end{tabular}
\end{center}
\end{footnotesize}

\section*{Appendix B: Validated cross sections}
We present validated cross sections for point SPS1a.  All sparticle
decays are turned off.  The non-default parameters used were chosen
mostly for simplicity, and to enable direct comparison with both the
\Py~6 ~\cite{Mrenna:1996hu,Sjostrand:2006za} and
\textsc{Xsusy} \cite{Bozzi:2007me} implementations: 

\begin{tabular}{ll}
PDF:pSet  = 8                     & (CTEQ6L1) \\
SigmaProcess:factorscale2 = 4     & ($\sqrt{ \hat s}  $) \\
SigmaProcess:renormScale2 = 4    & ($\sqrt{ \hat s} $) \\
SigmaProcess:alphaSvalue  = 0.1265 & \\
SigmaProcess:alphaSorder  = 1 &
\end{tabular}

\noindent
\begin{center}
\begin{tabular}{|l|rrrrrr|}
\hline
Process & Cross & Section & (fb) & & & \\
 & & & & & & \\

 \ttt{gg2squarkantisquark}  &  $\tilde d_L \tilde d_L^*$  &  $\tilde u_L \tilde u_L^*$
 &  $\tilde s_L \tilde s_L^*$  &  $\tilde b_1 \tilde b_1^*$  &
 $\tilde t_1 \tilde t_1^*$  & \\

          &     95.1  &     103.1  &      95.1  &    179.2  &    780.2  &  \\
 & & & & & & \\

 \ttt{qqbar2squarkantisquark}  &  $\tilde d_L \tilde d_L^*$  &  $\tilde u_L \tilde u_L^*$
 &  $\tilde d_L \tilde u_L^*$  &  $\tilde s_L \tilde s_L^*$  &
 $\tilde b_1 \tilde b_1^*$  &  $\tilde t_1 \tilde t_1^*$  \\

         &          59.9  &    89.6    &    64.6    &        30.8  &       48.7  & 154.3  \\
\ttt{onlyQCD}  &     63.9  &    97.4    &    87.6    &        30.7  &       48.3  & 153.5  \\
 & & & & & & \\

  \ttt{qq2squarksquark}  &  $\tilde d_L \tilde d_L$  &  $\tilde u_L \tilde u_L$  &
 $\tilde d_L \tilde u_L$  &  $\tilde s_L \tilde s_L$  &  $\tilde b_1
 \tilde b_1$  & \\
     &          130  &       459  &  765  &         5.11  &       1.06  & \\
 \ttt{onlyQCD} &     106  &       374  &  523  &         4.08  &       0.83  & \\

 & & & & & & \\
 \ttt{qg2squarkgluino} &  $\tilde g \tilde d_L$  &  $\tilde g \tilde u_L$  &
 $\tilde g \tilde s_L$  &  $\tilde g \tilde c_L$  &  $\tilde g \tilde b_1$  &  \\
       &            2.01  &            4.34  &           0.345  & 0.197  &     0.163  & \\
 & & & & & & \\

\ttt{gg2gluinogluino} & $\tilde g \tilde g$ &  & & & & \\
 &  0.142  & & & & & \\
 & & & & & & \\
 
\ttt{qqbar2gluinogluino} & $\tilde g \tilde g$ &  & & & & \\
&  2.97  & & & & &\\
& & & & & & \\
 \hline
\end{tabular}
\end{center}

\bibliographystyle{h-physrev3}
\bibliography{py8susy}

\end{document}